\documentstyle[epsfig,longtable]{aipproc}

\begin{document}
\title{The Standard Model Higgs in $\gamma \gamma$ Collisions}

\author{Michael Melles$^*$}
\address{$^*$Paul Scherrer Institute, Villigen CH-5232, Switzerland}

%\lefthead{LEFT head}
%\righthead{RIGHT head}
\maketitle

\begin{abstract}
For a Higgs boson below the $W^\pm$ threshold, the $\gamma \gamma$ collider
option of a future linear $e^+ e^-$ machine is compelling. In this case
one can measure the ``gold-plated'' loop induced
$\Gamma ( H \longrightarrow \gamma \gamma)$ partial width to high precision,
testing various extensions of the Standard Model. With recent progress in
the expected $\gamma \gamma$ luminosity at TESLA, we find that for
a Higgs of 115 GeV a statistical accuracy of the two photon partial width
of 1.4 \% is possible. The total width depends thus solely on the
accuracy of $BR(H \longrightarrow \gamma \gamma )$ and is of ${\cal O}
(10 \%)$.
\end{abstract}

The two photon Higgs width $\Gamma ( H \longrightarrow
\gamma \gamma)$, measured at the $\gamma \gamma$ Compton-backscattered option
of a future linear $e^\pm$ collider, is a very important physical
quantity \cite{bbc}.
In Ref. \cite{ddhi}
it was found that the MSSM and SM predictions can differ in the percentile
regime for large masses of the pseudoscalar Higgs $m_A$, depending mainly
on the chargino-masses.
The SM with two Higgs doublets (2HDM) and all other Higgs particles heavy
differs by about 10\% \cite{gko}.
At the PLC one measures the product $\Gamma ( H \longrightarrow \gamma \gamma)
\times
BR ( H \longrightarrow b \overline{b})$ and it is assumed that the branching ratio
can be measured in the $e^\pm$ mode via $BR ( H \longrightarrow b \overline{b}
)=
\frac{[\sigma (ZH) \times BR ( H \longrightarrow b \overline{b})]}{
\sigma (ZH)}$ with a 1 \% accuracy \cite{br}.
It was recently demonstrated in Ref. \cite{msk,m} that using conservative assumptions
an accuracy of 2\% is feasible for the two photon Higgs width
at a PLC.
There has been considerable progress in the theoretical understanding of the
BG to the intermediate mass Higgs boson decay into $b \overline{b}$ recently.
The Born cross section for the $J_z=0$ channel is suppressed by $\frac{m_q^2}{
s}$ relative to the $J_z=\pm2$ which means that by ensuring a high degree of
polarization of the incident photons
one can
{\it simultaneously} enhance the signal and suppress the background.
QCD radiative corrections can remove this suppression, however, and large
bremsstrahlung
and double logarithmic corrections need to be taken into account.
In Ref. \cite{jt} the exact one loop corrections to $\gamma \gamma
\longrightarrow q \overline{q}$ were calculated and the largest virtual correction
was contained in novel non-Sudakov double logarithms. For some choices of
the invariant mass cutoff $y_{cut}$ even a negative cross section was obtained
in this approximation. The authors of Ref. \cite{fkm} elucidated the physical
nature of the novel double logarithms and performed a two loop calculation
in the DL-approximation. The results restored positivity to the physical
cross section. In Ref. \cite{ms1}, three loop DL-results were presented which
revealed a factorization of Sudakov and non-Sudakov DL's and led to the
all orders resummation of all DL in form of a confluent hypergeometric function
$_2F_2$. The general form of the expression is $\sigma_{DL}=\sigma_{Born}(1+
{\mathcal F}_{DL}) \exp({\mathcal F}_{Sud})$.
In Ref. \cite{ms2} it was demonstrated that at least four loops on the
cross section level are required to achieve a converged DL result.
At this
point the scale of the QCD-coupling is still unrestrained and differs by
more than a factor of two in-between the physical scales of the problem,
$m_q$ and $m_H$. This uncertainty was removed in Ref. \cite{ms3} by introducing
a running coupling $\alpha_s ({\bf l^2_\perp})$ into each loop integration,
where $l_\perp$ denotes the perpendicular Sudakov loop momentum.
The effect of the RG-improvement
lead to
$\sigma^{RG}_{DL}=\sigma_{Born}(1+
{\mathcal F}^{RG}_{DL}) \exp({\mathcal F}^{RG}_{Sud})$.
The effective scale, defined simply as the one
used in the DL-approximation which gives a result close to the RG-improved
values, depends on the energy detector resolution
$\epsilon$, however in general is rather much closer to
$m_q$ than $m_H$ \cite{ms3}.
On the signal side, the relevant radiative corrections have long been known
up to NNL order in the SM \cite{dsbz,dsz} and are summarized including the MSSM
predictions in Ref. \cite{s}.
For our purposes the one loop corrections
to the two photon Higgs width are sufficient as the QCD corrections
are small in the SM.
The important point to make here and also the novel feature in this analysis is
that the branching ratio BR ($H \longrightarrow b \overline{b}$) is corrected
by the same RG-improved resummed QCD Sudakov form as the continuum heavy quark
 background
 \cite{msk}. This is necessary in order to employ the same two jet definition for
 the final state. Since we use the renormalization group improved
 massive Sudakov form factor ${\mathcal F}^{RG}_{Sud}$ of Ref. \cite{ms3}, we 
 prefer the
 Sterman-Weinberg jet definition \cite{sw}. 
 This is also necessitated by the fact that for three jet-topologies new
 DL corrections would enter which are not included in the background
 resummation of Ref. \cite{ms1}.
 We also use an all orders resummed running quark mass evaluated
 at the Higgs mass for $\Gamma ( H \longrightarrow b \overline{b})$. For
 the total Higgs width, we include the partial Higgs to $b \overline{b}, c
 \overline{c}, \tau^+\tau^-, WW^*, ZZ^*$ and $gg$ decay widths with all relevant
 radiative corrections.
 We begin with a few generic remarks concerning the uncertainties in our
 predictions. The signal process $\gamma \gamma \longrightarrow H \longrightarrow
 b \overline{b}$ is well understood and NNL calculations are available. The
 theoretical error is thus negligible \cite{s}.
 There are two contributions to
 the background process
 $\gamma \gamma \longrightarrow q \overline{q}$ which we neglect in this paper.
 Firstly, the so-called resolved photon contribution
 was found to be a small effect, e.g. \cite{jt}, especially since
 we want to reconstruct the Higgs mass from the final two-jet measurements and
 impose angular cuts in the forward region. In addition the good charm suppression
 also helps to suppress the resolved photon effects as they give the
 largest contribution.
 The second contribution we do not consider here results from the final state
 configuration where
 a soft quark is propagating down the beam pipe and the gluon and remaining
 quark form two hard back-to back-jets \cite{bkos}. We neglect this contribution
 here due to the expected excellent double b-tagging efficiency and the strong
 restrictions on the
 allowed acollinearity discussed below.
 A good measure of the remaining theoretical uncertainty in the continuum background
 is given by scanning it below and above the Higgs resonance. For precision extractions
 of $\Gamma ( H \longrightarrow \gamma \gamma )$ the exact functional form for resonant
 energies is still required, though.
 In terms of possible systematic errors, the most obvious effect comes from the
 theoretical uncertainty
 in the bottom mass determination. Recent QCD-sum rule analyses, however, reach below
 the 2\% level
 for $\overline{m}_b(\overline{m}_b)=4.17 \pm 0.05$ \cite{hg}
 including the effect
 of a massive charm \cite{m1}.
 For quantitative estimates of expected systematic experimental
 errors it is clearly too early to speculate at this point. The philosophy adopted 
 henceforth is that
 we assume that they can be neglected at the 1\% level and concentrate purely on the 
 statistical
 error.
 We focus here not on specific predictions for cross sections, but instead on the
 expected statistical accuracy of the intermediate mass Higgs signal at a PLC.
 As detailed in Refs. \cite{bbc}, due to the narrow Higgs width, the signal event
 rate is proportional to $N_S \sim \left. \frac{dL_{\gamma \gamma}}{dw}
 \right|_{m_H}$, while the BG is proportional to $L_{\gamma \gamma}$.
 To quantify this, we take the {\it updated} design parameters of the proposed TESLA
 linear collider  \cite{t,tp},
 which correspond to an integrated peak
 $\gamma \gamma$-luminosity,
  of 40 fb$^{-1}$ for the low energy running of the
  Compton collider.
  This corresponds to an integrated geometrical luminosity $L_{e^\pm}$
  of 400 fb$^{-1}$ and a
  conversion coefficient $k^2=0.4$ and a $10^7$ sec. run at TESLA.
  The polarizations of the incident electron beams and the
  laser photons are chosen such that the product of the helicities $\lambda_e
  \lambda_{\gamma} = -1$.
  This ensures high monochromaticity and polarization of the photon beams
  \cite{t,tp}.
  Within this scenario a typical resolution of the Higgs mass is about 10~GeV, so
  that for comparison with
  the background process $BG \equiv \gamma \gamma \longrightarrow q \overline{q}$
  one can use \cite{bbc}
  $\frac{L_{\gamma \gamma}}{10\; {\rm GeV}} = \left.
  \frac{d L_{\gamma \gamma}}{dw}\right|_{m_H}$
  with $\left. \frac{d L_{\gamma \gamma}}{dw} \right|_{m_H}=$1.7 fb$^{-1}$/GeV.
  The number of background events is then given by
  $N_{BG} = L_{\gamma \gamma} \sigma_{BG}$.
  In Ref. \cite{msk} it was demonstrated that in order to achieve a large enough
  data sample, a central thrust angle cut $| \cos \theta | < 0.7$ is
  advantageous and is adopted here. We also assume a (realistic) 70\%
  double b-tagging efficiency. For the charm rejection rate
  it seems possible to assume 
  improvement
  from a better single point resolution, thinner detector modules and
  moving the vertex detectors closer to the beam-line \cite{ba}.
  With these results in hand we keep $| \cos \theta | < 0.7$ fixed
  and furthermore assume the $\overline{c} c$ misidentification rate of 1\%
  \cite{msk}. We vary
  the cone angle $\delta$ between narrow ($10^o$), medium ($20^o$) and large
  ($30^o$) cone sizes for both $\epsilon=0.1$ and $\epsilon=0.05$.
  The analysis in Ref. \cite{msk} using not the updated machine parameters
  demonstrated that for a Higgs
  boson with $m_H < 130$ GeV, the statistical precision
  can be below the 2\% level after collecting one
  year of data. 
  The good charm misidentification rate is important
  for $\sqrt{N_{tot}}/N_S$.
  The level of accuracy was confirmed in Ref. \cite{js} with an independent MC-simulation.
  The previous analyses, however, did not take into account the
  new luminosity assumptions. These lead to an improvement for the statistical
  significance of a factor $\frac{1}{\sqrt{3.3}}\approx 0.55$.
  Together with the expected uncertainty of 1\% from the $e^+e^-$ mode determination
  of BR$(H \longrightarrow \overline{b} b)$ \cite{ba},
  we conclude that a measurement of the partial width
  $\Gamma (H \longrightarrow \gamma \gamma)$ of 1.4\% precision
  level\footnote{We assume four years of running to determine BR$(H \longrightarrow \overline{b} b)$
  at 1\%,
  uncorrelated error progression and negligible
  systematic errors.}
  is feasible for a Higgs of 115 GeV from a purely statistical point of view.
  In the entire MSSM mass range the precision below the 2\% level is achievable.
  This level of accuracy could significantly enhance the kinematical reach
  of the MSSM parameter space
  in the large pseudoscalar mass limit or the 2HDM and thus open up a window for
  physics beyond the Standard Model.
  For the total Higgs width, the expected precision is dominated by the
  branching ratio BR($H \longrightarrow \gamma \gamma$), which can be determined
  at the 10 \% level \cite{sch}.
  In summary, using realistic and optimized machine and detector design parameters, we
  conclude that the Compton collider option at a future linear collider
  can considerably extend our ability to discriminate between the SM and
  MSSM or 2HDM scenarios.
  \section*{Acknowledgments}
  I would like to thank my collaborators W.J.~Stirling and V.A.~Khoze for their 
  contributions to
  the results presented here.

\end{document}